# Optical identification using physical unclonable functions [Invited]

Pantea Nadimi Goki,[1,2,*] Stella Civelli,[2,3] Emanuele Parente,[2] Roberto Caldelli,[4,5] Thomas Teferi Mulugeta,[2] Nicola Sambo,[2] Marco Secondini,[2] Luca Potì[1,5]

[1]*Photonic Networks and Technologies Laboratory, CNIT, Via G. Moruzzi 1, 56124, Pisa, Italy*
[2]*TeCIP Institute, Scuola Superiore Sant'Anna, Via G. Moruzzi 1, 56124, Pisa, Italy*
[3]*CNR-IEIIT, Via Caruso 16, 56122, Pisa, Italy*
[4]*CNIT Florence Research Unit, Viale Morgagni 65, 50134, Florence, Italy*
[5]*Universitas Mercatorum, Piazza Mattei 10, 00186, Roma, Italy*
*\*Corresponding author: pantea.nadimigoki@santannapisa.it*

**In this work, the concept of optical identification (OI) based on physical unclonable function is introduced for the first time, to our knowledge, in optical communication systems and networks. The OI assigns an optical fingerprint and the corresponding digital representation to each sub-system of the network and estimates its reliability in different measures. We highlight the large potential applications of OI as a physical layer approach for security, identification, authentication, and monitoring purposes. To identify most of the sub-systems of a network, we propose to use the Rayleigh backscattering pattern, which is an optical physical unclonable function and allows to achieve OI with a simple procedure and without additional devices. The application of OI to fiber and path identification in a network, and to the authentication of the users in a quantum key distribution system are described.**

## 1. INTRODUCTION

The rapid growth of global communication networks around the globe requires optimal network security protocols.

Each layer of the open systems interconnection (OSI), which describes how different layers communicate in a network, contributes to the overall security of the network, which includes secure communication, authentication, identification, and monitoring. Figure 1 depicts the security protocols that can be applied to the layers of the OSI, briefly described below. The application layer, which is the layer with which most of the users interact, may include end-to-end cryptography, e.g., Outlook, and WhatsApp messages are encrypted to be recognized just by users. Also, the presentation and session layers, which are responsible for syntax processing and creating communication channels between devices, respectively, may profit from data cryptography. The transport layer, which is responsible for the transmission of data across network connections, may use secure sockets layer (SSL) or transport layer security (TLS) protocols that include authentication between parties, data integrity, and digital signature. The network layer, which handles the routing of the data, is responsible for security at the network level and uses functions such as packet authentication, cryptography, and integrity, e.g., Internet protocol security (IPsec). The data link layer uses admission controls to check and guarantee the proposed connection, for example, wireless systems developed Wi-Fi protected access (WPA). Concerning the physical layer, usually, security is not implemented because establishing optimal security protocols at this level is still an open worldwide problem. Although the upper layers are liable to security and confidentiality, implementing a security protocol on the physical layer could significantly enhance the network's security. Potential attacks that target the physical layer are included tampering (which introduces fake nodes), jamming (which introduces harmful signals in the network), side-channel attacks (when the adversary gets physical access to the device), physical infrastructure attacks, and eavesdropping. Hence, physical layer security (PLS) is a crucial element that can enhance the overall security of the networks. Although PLS cannot prevent such attacks, however, it might detect such attacks and warn the users. To establish PLS several methods have been proposed and studied. The very first technique based on information theoretic characterizations of secrecy for PLS is the Wyner technique which is defined by the wiretap channel model [1]. The Wyner technique limits the information to an eavesdropper by using the channel capacity difference between a target receiver and an eavesdropper, defining positive secrecy capacity only if the target receiver has better signal-to-noise ratio (SNR) than the eavesdropper, which makes this technique unsecured. An adversary with high-performance devices can receive higher SNR than the target receiver. The eavesdropper attack may be neutralized, by transmitting the artificial noise to reduce its channel capacity, only if the attacker's position is known [2]. Since then, various research has been done in this area, and most of them that guarantee security is based on suited encoding and complicated modulation schemes [3,4].

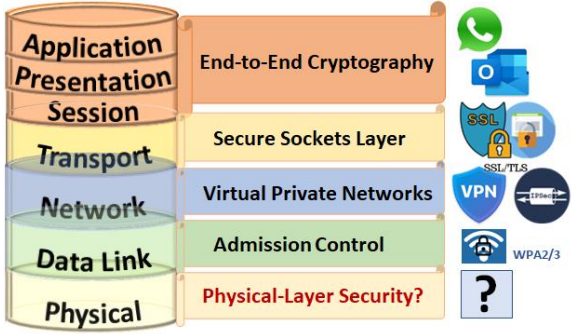

Fig. 1. Security along Open system interconnection (OSI) levels. The target of our proposed method in this paper is to implement a novel ID technique for physical layer security.

The PLS techniques based on computational cryptography rely on computational hardness but are vulnerable to digital attacks (e.g., Brut force attack). Quantum key distribution (QKD) [5,6] provides intrinsic security, but (i) is not cost-effective, (ii) is hard to be implemented, and (iii) relies on user authentication usually performed with classical techniques. Also, the PLS based on keys generated by digital signal processing (DSP) [7] is vulnerable to digital attacks. Recently, an approach based on optical steganography was proposed to hide messages below the noise level [8,9]. However, this technique cannot detect the presence of an eavesdropper and is vulnerable to adversaries who know the technique [10,11]. PLS-based optical chaos communication [12,13] requires high-level synchronization between the transmitter and the receiver [14] and its security can be broken in some scenarios [15].

A new approach to PLS is based on the material's physical features, defined by physical unclonable functions (PUFs), in which a physical device provides unique output for a given input [16]. The security of this method relies on the intrinsic unclonability of the PUF [17-19], and, therefore, is able to overcome the disadvantages of computational cryptography. Optical PUFs (OPUFs), PUF defined in the optical domain, have been recently studied [20-23]. Even though OPUFs have been investigated for secure cryptography key generation, they were never employed in a real system as a practical security solution [24].

The concept of optical fingerprint has been recently proposed to exploit the inherent characteristics of some physical devices [25-27]. In these works, classification and identification are obtained in limited scenarios, through training processes using specific equipment, and without strong reliability in terms of unclonability and security.

Despite all the effort aforementioned to implement PLS, it is still an open problem. Indeed, optical fibers, which constitute the larger part of the physical layer, are distributed around the globe and are vulnerable to adversarial attacks, whose capabilities are growing day by day thanks to the use of high-performance devices or exploiting intensive machine learning algorithms. In this paper, we propose a novel method to ensure network security: optical identification (OI). The OI is based on the Rayleigh backscattering pattern (RBP) extracted from an optical fiber, which is a strong OPUF [28]. The proposed method can be used for communication security, authentication, identification, and monitoring, both in point-to-point communication and optical networks. In particular, OI protects sub-systems' communication since it allows their physical identification. Indeed, any physical attack which affects fibers, fiber connections, or inserts optical devices (e.g., optical couplers/splitters) changes the system RBP, thus changing the system signature. For instance, spoofing, tampering, jamming [29], and eavesdropping attacks [30] modify the system signature and reveals the presence of an imposter.

This manuscript is organized as follows. In Section 2, we introduce the concept of optical identification and its security validation. In Section 3, we introduce the concept of optical physical unclonable function, and we describe the Rayleigh backscattering. Next, Section 4 describes some potential applications of the OI concepts, and Section 5 provides some examples. Finally, Section 6 draws conclusions.

## 2. OPTICAL IDENTIFICATION

### A. Concept

Not only humans but also physical elements have their own fingerprints. In general, the fingerprint (equivalently, signature) of a device, system, or sub-system, denoted below as ID, is related to its physical characteristics and caused by imperfections in the manufacturing process. In this manuscript, we propose to exploit the inherent characteristics of fiber network systems to produce an unclonable fingerprint to be used for security purposes, a concept that has not been exploited in communication systems and networks. We refer to this concept as optical identification (OI) and we briefly describe it in the following.

A point-to-point scenario, sketched in Figure 2(a), is made of three sub-systems: transmitter, channel, and receiver. Let us denote the signatures of these three sub-systems as $ID_{TX}$, $ID_{Ch}$, and $ID_{RX}$, respectively. In this case, three possible security approaches may be envisaged: (i) the transmitter reads $ID_{Ch}$ and $ID_{RX}$ to be sure that the information passes the expected channel and reaches the expected receiver, (ii) the receiver reads $ID_{TX}$ and $ID_{Ch}$ to be sure who is the sender and which is the physical path, (iii) the transmitter reads $ID_{RX}$ and the receiver acquires $ID_{TX}$ so that both know to whom they are talking (they can also acquire $ID_{Ch}$ to check the path).

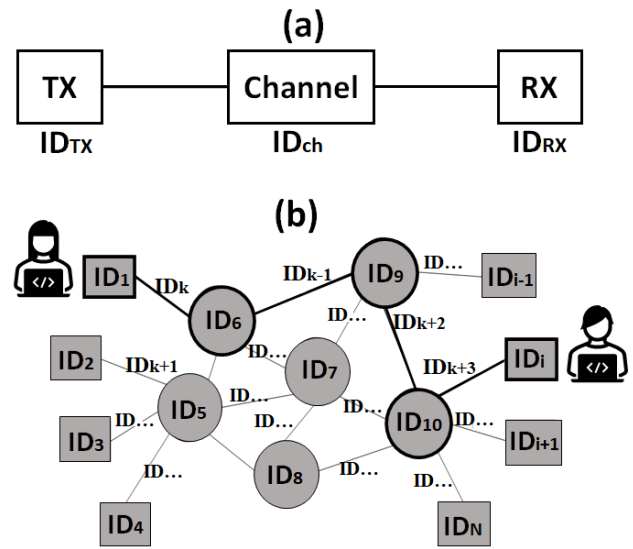

Fig 2. (a) A point-to-point communications system. (b) Example of network architecture.

In a network scenario with $N$ sub-systems, sketched in Figure 2(b), each sub-system may be identified by its signature, labeled as $ID_i$ where $1 \leq i \leq N$ represents the $i$-th sub-system. In an optical network, sub-systems may include transceivers, optical fibers, optical nodes, filters, optical cross-connects, reconfigurable optical add/drop multiplexers, etc.

The ID of each sub-system, both in a point-to-point scenario or the network scenario, can be generated and stored in a database. Each sub-system is identified by comparing its ID with the corresponding stored copy, and correct identification is characterized by probability $p_i$. In a network scenario, any sub-system may be able to identify another sub-system in the network and, eventually, validate the whole path. Let us consider the bold path in Figure 2(b). If the sub-system with $ID_1$ is able to evaluate the probability of each path and these paths are independent, each sub-system can be independently interrogated, and the probability of the whole path becomes:

$$p = \prod_{i \epsilon \{path\}} p_i \qquad (1)$$

The above model allows us not only to validate each sub-system and the path but also to identify whether any changes occurred in the path and where. However, the acquisition of the signature of the sub-system is not independent. Consequently, a more complex model based on the specific technique used for identification must be developed.

**B. Security validation**

The ID of each sub-system can be represented by a vector of bits, the digital representation of the signature. For the sake of brevity, we will refer to this as digital signature. How this digital signature is obtained depends on the specific implementation, and some examples are given in the next sections.

It is important to underline that, even if the physical signature is unclonable, its digital representation loses this property. As a consequence, is essential to be able to assess the accuracy and strength of a signature generation method. Below, we describe how to perform the identification of a binary signature, and how to assess its strength.

Let us assume there are two users U and V, in a point-to-point communication, each with their digital signature with $N$ bits. To compare two signatures, we use the inter-Hamming distance (HD), which counts the number of different bits among the two signatures.

Assuming that the bits of each signature are independent and identically distributed and that $p$ is the probability of having different bits in the two IDs, the HD of two signatures is distributed as a binomial distribution with $N$ trials and mean value $Np$. This means that when two IDs are independently generated, $p = 0.5$ and the mean is $N/2$. Conversely, when two IDs are not dissimilar (which happens when they represent the same sub-system) a few bits should be flipped to obtain one from the other, i.e., $p$ and the HDs are small. Consequently, the decision rule is: if HD is below a certain threshold $t$, we assume that the two IDs represent the same user; conversely, if the HD is larger than $t$, we assume that the IDs belong to different users.

This concept is illustrated in Figure 3, which reports the probability of the HD between the IDs of the users U and V with the stored ID of the user U. The figure serves only for illustration purposes and, therefore, we do not report the scale on the axis or the system setup. The figure shows that the mean of the HD for U, denoted as $M_U$, is much lower than the one for V, denoted as $M_V$. The threshold $t$ can be defined as

$$t = \gamma M_V + (1-\gamma) M_U \qquad (2)$$

where $0 \leq \gamma \leq 1$.

The success or failure of the procedure depends not only on intrinsic physical limitations and inaccuracies (e.g., the amount of noise in the RBP acquisition) but also on the post-processing method i.e., signature definition and decision rule (e.g., the number of bits). The procedure fails when a false negative or a false positive occurs. On the one hand, a false negative "U rejected" occurs when U is the user, but the procedure fails, and he is rejected (the HD is larger than $t$). This is a matter of identification, which can be partly mitigated by repeating the identification protocol several times. On the other hand, a false positive "V accepted" occurs when the user is V (different from U), but the procedure fails, and V is identified as U (the HD is smaller than $t$). This is a matter of security and authentication. In general, while is it desirable to minimize both the probability of false negative and false positive, one can tailor $t$ to the system requirements: if $t$ decreases, the security improves (the probability of false positive decreases) at the expense of identification capabilities (the probability of false negative increases). The probability of failure (the sum of the probability of false positive and false negative) can be evaluated (i) estimating by simulations the mean of the HD of the right user $M_U$ (two signatures of U) and of the wrong user $M_V$ (the signature of U and the signature of V) (ii) considering the two binomial distributions with $N$ trial, and probability of success $M_U/N$ and $M_V/N$, respectively, and (iii) estimating the probability of false positive as:

$$\sum_{k=0}^{HD} \binom{n}{k} \left(\frac{M_V}{N}\right)^k \left(1 - \frac{M_V}{N}\right)^{N-k} \qquad (3)$$

and the probability of a false negative as:

$$\sum_{k=HD}^{\infty} \binom{n}{k} \left(\frac{M_U}{N}\right)^k \left(1 - \frac{M_U}{N}\right)^{N-k} \qquad (4)$$

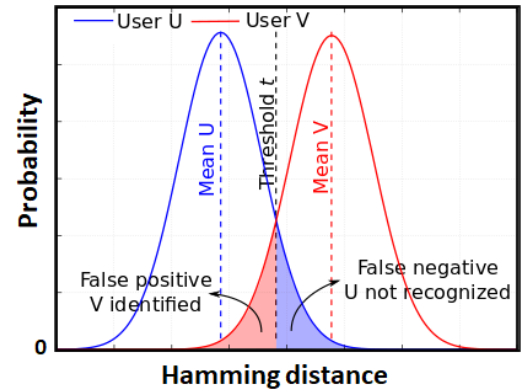

Fig. 3. Probability of HD of two users. The false positive and false negative regions are highlighted, depending on the decision rule.

Depending on the scenario and type of signature, it is straightforward to use quick response (QR) codes to represent signatures (simply binary matrices). In this case, the robustness of the method also can be evaluated by considering the inner HD and the intra-HD, as in [31]. These two are useful to demonstrate the robustness of the signature against a digital forecasting attack and to demonstrate ID reproducibility. Inner HD is the HD of a pair of 1D segments in the QR code that should not be too smaller or bigger than half of the size of the segment length to be robust against digital attacks.

Intra-HD is the HD of several repeated measures of the same sub-system, and should be low to indicate the reproducibility of the ID.

## 3. OPTICAL PUF

### A. Concept

A physical unclonable function (PUF) is, by definition, a function that, under specific circumstances and for a given input (referred to as *challenge*) provides a unique output (*response*) that results to be unclonable [32]. The uniqueness of the signature is a physical characteristic of the PUF, usually due to the imperfections of the manufacturing process. A PUF can be generated by physical objects, like the communication and network sub-systems, e.g., sensors, integrated circuits, and hardware in general. In a nutshell, PUFs provide a signature, or fingerprint [33], of physical devices, which can be used for security applications.

A PUF is a black-box function $F(\cdot)$ which provides a unique output, the *response* $R = F(C)$, given as input the *challenge* $C$. We refer to the pair $C$ and $R$ as the challenge-response pair (CRP). Figure 4 sketches how two different PUFs A and B (with functions $F_A(\cdot)$ and $F_B(\cdot)$) respond to two different challenges $C_1$ and $C_2$. On the one hand, given two different challenges $C_1$ and $C_2$, the responses of the same PUF A, $F_A(C_1)$ and $F_A(C_2)$, are different. On the other hand, the same challenge $C_2$ provides two different responses $F_A(C_2)$ and $F_B(C_2)$, when using two different PUFs.

Soft PUFs are used for PUFs with limited number of challenges, while strong PUFs are used for PUFs with a large number of challenges. In the latter case, generally, the complete determination of the CRP is not possible in a feasible way. The property of uniqueness is defined by means of the inter-Hamming distance of the outputs, that is how different are the responses of distinct PUFs. The reliability of a PUF is the ability to provide the same response for a given challenge; this is measured by the intra- Hamming distance, which is the HD among two responses to the same challenge and should ideally be equal to zero; while steadiness indicates the variability of the response due to changes in the circumstances e.g., temperature, power supply or aging effect [34].

Different kinds of PUFs exist, some of which are described in the following. The system presented in [35] is one of the first examples of a strong optical PUF. In this case, an input laser beam is directed towards a stationary scattering medium and then the speckle output pattern is recorded. The laser XY location and its polarization constitute the challenge while the response is the associated speckle pattern. Such a pattern is strongly dependent on the input location/polarization due to the fact that multiple scattering events can occur inside the scattering medium. Conversely, the power-on state of a static random access memory (SRAM) is a soft PUF. In fact, though an SRAM cell is symmetric, manufacturing anomalies can induce a tendency toward a logical "1" or "0" when the power is switched on. This variability is random across the entire SRAM and can determine a univocal fingerprint. Another interesting example of soft PUF is obtained in the case of digital image (video) acquisition. When a photo is acquired, the camera sensor, which is composed of a two-dimensional array of charge-coupled devices (CCDs), is hit by light photons and this energy is then converted into electron charges. Due to manufacturing imperfections, each cell of such a silicon sensor differently answers to a uniform incoming light. Consequently, this results in the superimposition, in each content it takes (images and/or videos), of a systematic noise, named photo response non-uniformity noise (PRNU) [36]. The PRNU is not perceivable and does not degrade the visual quality of the acquired contents, but it constitutes a fingerprint that is embedded within the image pixels and can be used for source identification.

Overall, it is evident that PUFs have strong characteristics that can be used for optical identification (OI). For example, PUFs can be used for authentication purposes, by storing a database of CRPs $(C_i, R_i)$. Furthermore, PUFs can be used as a cryptographic root key for a device, such that key injection is not required, and the key cannot be copied and does not need to be stored but is simply recovered from the device when necessary.

However, despite the advantages, PUFs are anyway prone to security issues and should be carefully tackled in relation to the application scenario.

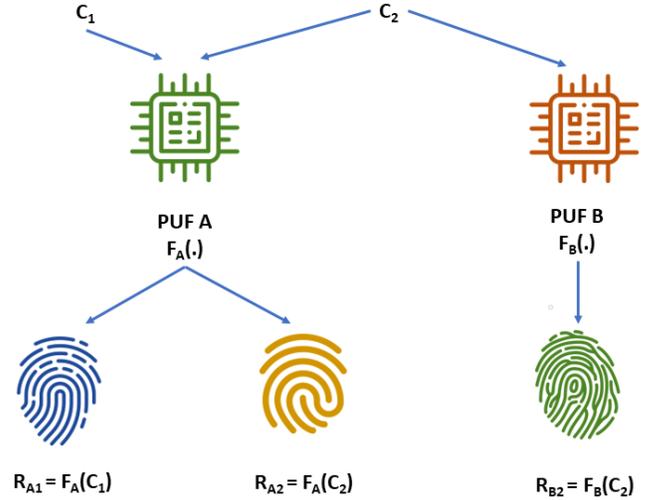

Fig. 4 Challenge-response pairs (CRPs).

### B. Rayleigh backscattering as an OPUF

The Rayleigh backscattering that occurs when stimulating optical fibers with propagating light is an OPUF response, due to the random density fluctuations caused by the fabrication process [28]. Therefore, we propose to use the Rayleigh backscattering pattern (RBP) as a signature of the optical fiber, which allows us not only to identify the fiber link but any optical and opto-electronic sub-systems through their pigtail.

RBP acquisition can be done using the optical frequency domain reflectometry (OFDR) technique with sub-millimeter-level spatial resolution [36-40]. Furthermore, it is known that due to the sensing capability of OFDR, the RBP is robust against changes in temperature or strain [28].

In this work, we consider the coherent OFDR (C-OFDR) since it allows us to increase the sensitivity and resolution [41,44].

C-OFDR is implemented as follows. The light from a continuous wave (CW) laser with amplitude $E_0$, whose frequency is linearly swept in time with sweep rate $\gamma$, propagates into the fiber under test (FUT). The RBP is the photocurrent obtained after self-coherent balance detection, as sketched in Figure 5, and can be modeled as

$$I(t) = E_0^2 \sum_{k=1}^{n} \sqrt{R_k} \cos(2\pi\gamma t\tau_k) \quad (5)$$

when there are $n$ reflection points with reflectivity $R_k$ and roundtrip time $\tau_k$ [28,44]. For the sake of simplicity, a responsivity of 1 is assumed. The photocurrent in Eq. (5) is acquired with an analog-to-digital converter (ADC) and can be used for optical identification purposes. For example, using a single-bit ADC (or digitally emulating it), one can directly use the security validation method in Section 2B.

We remark that the RBP in (5) is the response of the OPUF, which depends on the challenge applied. Consequently, any fiber possesses several different fingerprints, each depending on the applied challenge. This is a key aspect for optical identification, since it protects from the reproduction of the fingerprint without knowledge of the challenge, as specified below.

The major feature provided by the RBP is its unclonability, which intrinsically protects against spoofing attacks. Indeed, spoofing, which consists in faking the ID and identifying it as another sub-system, requires either replacing legitimate fiber or generating a fake signal. On the one hand, if the legitimate fiber is replaced, it causes a different RBP (even if the adversary uses the correct challenge) because the RBP is a PUF response and is unique for each fiber, and the signatures do not match. On the other hand, RBP is measured through a self-coherent receiver where the same laser is used as the source and local oscillator. Therefore, since laser phase and wavelength cannot be predicted or fully measured, RBP reproduction is avoided.

It is important to note that fiber fingerprint (signature) defines based on the specific challenge that applies to the fiber. Consequently, any fiber can possess several different fingerprints that define implementing various challenges. That is a significant fact that causes a high-security level for the generated signature. Since fingerprint is dependent on the challenge, an adversary that has access to the fiber cannot generate the signature without applying the related challenge. Accordingly, the response (signature) must be stored in the database along with the related challenge and constitute the (CRP) database for the identification approach, as is illustrated in Figure 12-13. It is worth noting that applying one challenge on different fibers results in various responses, as explained in section 3. A, every single fiber has its unique response.

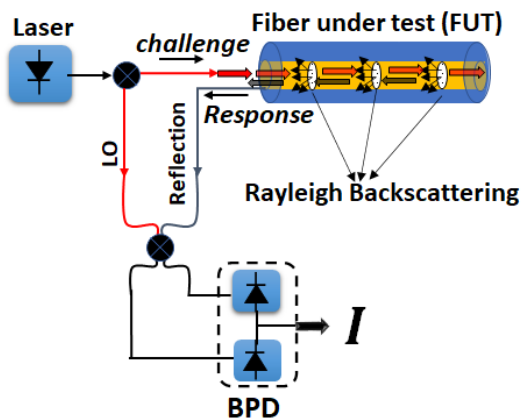

Fig. 5. RBP acquisition with C-OFDR. LO: local oscillator. BPD: Balanced photodetector.

## 4. POTENTIAL APPLICATIONS

### A. Physical layer security in classical communications

In this section, the possibilities and challenges of optical identification (OI) – or more in general of fingerprints – are discussed. The RBP of an optical fiber is an OPUF, and, therefore, each fiber can be characterized by a fingerprint. This fingerprint can be obtained, for example, offline before the fiber installation, having access to an end of the fiber. Let us assume that the operator maintains a database with the fingerprints of all fibers in the network, thus having awareness of all installed fibers. In this way, unauthorized access to the network can be revealed. We can assume that an operator may be aware of the fingerprint of its fibers over an infrastructure (e.g., before the installation of the infrastructure). Thus, assuming that each fiber is known, the operator can periodically perform checks (e.g., from the node ports) and reveal if some intrusion has been performed (e.g., some fiber has been replaced with some other fiber). The intrusion can be then revealed through the proposed method when an unknown fingerprint is detected

In particular, potential applications are the identification of tampering (consisting of making fake nodes) or jamming (introducing harmful signals in the network). Indeed, fibers – e.g., attached to an edge port – can be checked and "authorized" periodically or on-demand according to management policies. This technique can be used to identify switch ports in case of an optical bypass of the switches, e.g., aggregate switch identification by the core switch in colocation data center scenarios. It is worth mentioning that the identification strategies could be different for different systems and networks [45] depending on distance and the devices in the system (e.g., router).

Finally, a relevant issue related to amplifiers is here discussed. Amplifiers typically include isolators that limit the propagation of the signal-stimulating RBP. Thus, because of isolators, it is not possible to measure the fingerprint of a concatenation of fiber spans. Such an issue mainly impacts applications to backbone or metro networks, where amplifiers are typically employed. This may prevent an operator to limit the number of monitoring points for authentication. Indeed, if estimating the fingerprint of a concatenation of fibers from the knowledge of each fiber fingerprint would be possible, this could be checked against its measurement on the field identifying an intrusion in a point in between, without checking span by span. Differently, in a short-reach scenario, such as intra-data center or PONs where amplifiers are typically not needed, estimating the fingerprint of a concatenation of fibers can be very useful. Thus, this can pave the way for the study of models estimating the fingerprint of concatenated fibers. At the moment only preliminary works have been done on fingerprint concatenation [45,46] and more detailed studies are needed.

Another relevant challenge can be the automation of authentication and identification procedures relying on OPUF with the proposal of properly designed protocols, initiating the authentication procedure, disseminating measured fingerprints, correlating such measurements, and sending possible alarms when detecting intrusions.

### B. User authentication in quantum system

Since the term PUF was coined by Pappu et al. in 2001 [35], these objects gained a lot of interest and started to be used for a wide series of different security purposes such as identification and authentication, with applications in tamper evidence, anti-counterfeiting, etc. They have recently been adopted to face the authentication problem in quantum key distribution systems [47]).

A generic QKD protocol is able to offer information-theoretic security (ITS) and its aim is to allow two users to establish a common secret key despite the presence of powerful adversaries. In order to succeed in this, two users employ a quantum channel that is thought to be open to possible tampering by an eavesdropper and a classical one which, instead, needs to be authenticated. Under these assumptions, an attacker can manipulate the raw key created via the exchange of the quantum states and only listen to the conversation over the classical channel. Here, with respect to classical protocols (symmetric and asymmetric classic cryptographic schemes), the laws of quantum mechanics provide the possibility to estimate a possible eavesdropper's intervention and the potential amount of information in her hands, so that the protocol can be eventually stopped [48]. This evaluation happens during the post-processing stage, performed along the classical channel: in this scenario, it appears of absolute importance a proper authentication of the classical channel, as each of the two legitimate parties of the conversation needs to rely on the other's party true identity so to prevent a possible man-in-the-middle attack. In fact, a malevolent party, say Eve, can connect her QKD devices to the loose ends of the channels in order to hide her presence (Figure 6) so that she pretends to be Bob to Alice and Alice to Bob, modifying any message sent from Alice to Bob or vice versa.

The tool for this authentication job is the so-called message authentication code (MAC), for whose realization the Wegman-Carter authentication scheme and variations thereof are the most implemented methods to provide ITS [49]. Anyway, this kind of authentication requires a pre-shared key, which is usually considered the main drawback of QKD protocols; moreover, the need for a pre-shared secret key complicates considerably the design of large full-mesh QKD networks, as the number of keys has a quadratic grown with the number of users participating. In the past, different efforts were made to decrease the length of the pre-shared key in existing QKD protocols and to make easier their distribution and management [50].

A possible solution to this problem is the integration of the PUFs at the endpoints of the classical channel of a QKD apparatus, using their response as the tag generation required for the classical channel authentication. This tag is characterized by internal random disorder because the response to a given challenge reflects the internal disorder of the device: in this way, the response of a PUF can play the role of a fingerprint [51,52].

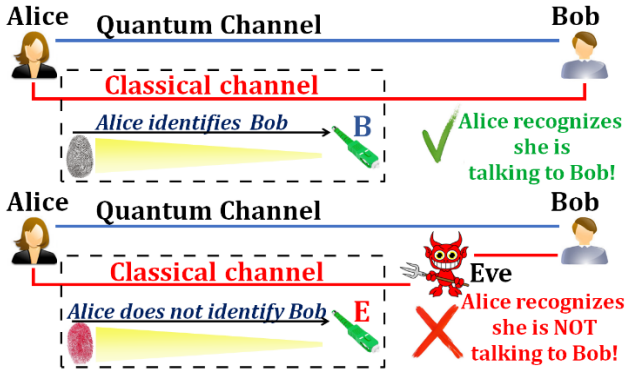

Fig.6. Schematic representation of the proposed authentication protocol: Alice can recognize whom she is talking with thanks to the Rayleigh backscattered light from the other's party pigtail, as this provides a unique and unclonable fingerprint of her interlocutor.

Due to the unpredictable feature of PUFs, their use adds another layer of security in a QKD apparatus. The existing PUFs collections are so extensive that every class of them is characterized by features that can be useful concerning some applications. While Nikolopoulos [52] has considered the PUF tag, which is fabricated and attached to each QKD box (sender or receiver), we introduce here an OPUF, which is already present in the QKD box without manufacturing requirements and is an inherent feature of the QKD box, that eliminates such PUF tag disadvantages as tag scratching or stealing, or copying by adversaries, who have access to the PUF tag. Considering that each QKD box includes the optical fiber (fiber optic transceiver pigtails), each QKD box can be uniquely identified by its ID generated by our proposed OPUF-based identification model, which is the RBP of its fiber optic pigtails. In this way, each QKD box carries the tag inside itself, i.e., its ID, which is hidden from the rivals, and just only the one that can measure it can observe the ID, whereas the external tag ID can be observed and copied by the adversaries. In summary, our proposed model is based on strong OPUF and seems a promising candidate for the authentication problem in the QKD system, which is also compatible with QKD infrastructures.

## 5. PRACTICAL OI APPLICATIONS IN OPTICAL COMMUNICATION SYSTEMS AND NETWORKS

In optical communication systems or networks, OI can be implemented through the measurement of the RBP of the device pigtail or of the fiber link. Below we describe some implementation examples.

### A. Sub-system identification

Let us consider an optical sub-system having its own fiber pigtail whose length is generally in the order of 50-100 cm, which we use for RBP measure.

We consider a simulation scenario with a 0.5m fiber pigtail, and we measure the RBP with the C-OFDR with a sweep time of 0.5s. The analog signal is digitized using a single-bit analog to digital converter (ADC) with $N = 4000$ samples.

For the sake of comparison with [28], 96 additional bits are added, and the signature is converted into a two-dimensional (2D) $64 \times 64$ matrix, graphically represented as a QR code in Figure 7, with $n = 64$ (the length of each row). Note that an additional level of security can be added in the 96 bits, in the form of a shared key.

Firstly, we investigate the robustness of the signature against brute force trials (BFT) by evaluating its inner-Hamming distance. Comparing all possible combinations of the 64 rows (which provides 2016= $\binom{64}{2}$ pairs), we obtain 2016 HD values, ranging from 0 to $n = 64$. If the inner HD is too small or too large, it means that the rows are correlated, and one can be easily obtained from the other one. Conversely, it is desirable to have inner HD in average close to $n/2 = 32$. In our case, the histogram of the inner HD, shown in Figure 7(c), has a mean value of 31.9, very close to the optimal value.

Next, to assess the reproducibility and uniqueness of the signature, we consider the intra-Hamming and inter-Hamming distances, respectively. We generated 100 different signatures of the same sub-system, adding some random white Gaussian noise (in a practical system, repeated measurements are subject to noise). It is desirable that the HD of these signatures, the intra HD, is small, which indicates that the signature match is more probable. We also generated 100 different signatures representing 100 subsystems. Conversely, it is desirable that the HD among the signatures of different subsystems, the inter HD, is close to $n^2/2$, to ensure that it is difficult to match by mistake. Figure 8 (a) and (b) show the intra and inter-HD obtained by our simulations. The first has a mean value of 144.06, while the second a mean value of 1994.08, both showing excellent values.

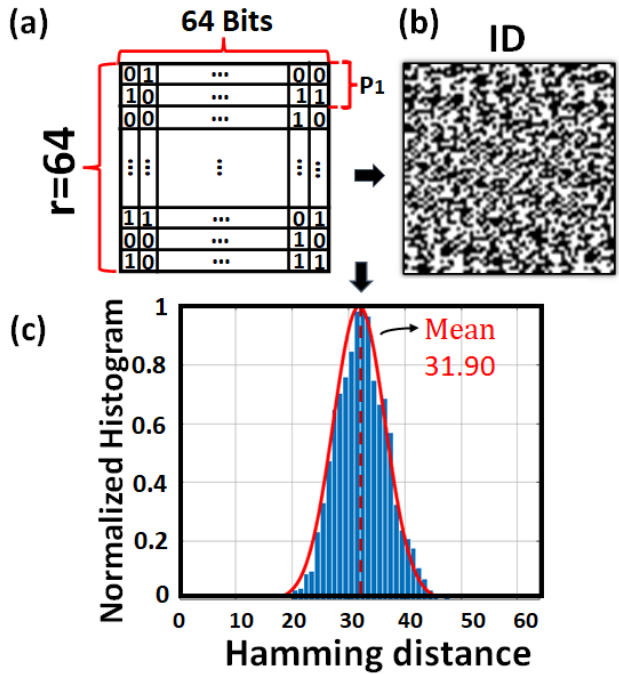

Fig. 7. Validation of the ID robustness against BFT attacks. (a) 2D binary ID data consists of $64 \times 64$ bits. (b) The QR code. (c) Histogram (and Gaussian fit) of the inner-Hamming distance between the 2016 pairs of its 64 rows.

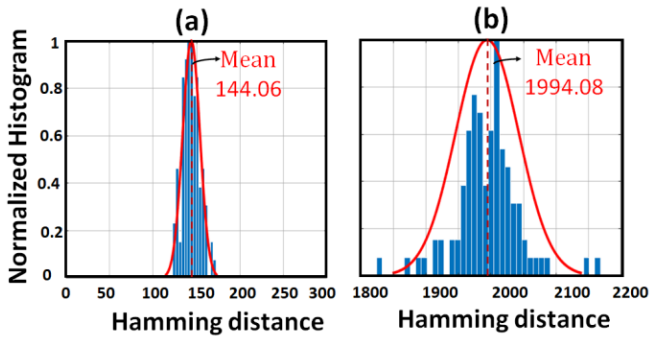

Fig. 8. Histogram of simulated 100 IDs for repeating the testing. (a) Intra-Hamming distance of one sub-system with 100 times measurements, shows ID reproducibility. (b) Inter-Hamming distance of 100 different IDs stored in the CRP database and one ID, which was not stored in the library, shows ID uniqueness.

We show in Figure 9 how the probability of false negative and false positive changes when the threshold changes. When $\gamma$, as defined in equation (2), approaches zero (one, respectively), the probability of false negative and false positive becomes 0.5 (minimum, respectively), while the probability of false positive becomes minimum (0.5, respectively). For $\gamma = 0.5$ and $SNR = 0$ dB, the probability of false identification is in the order of $10^{-60}$.

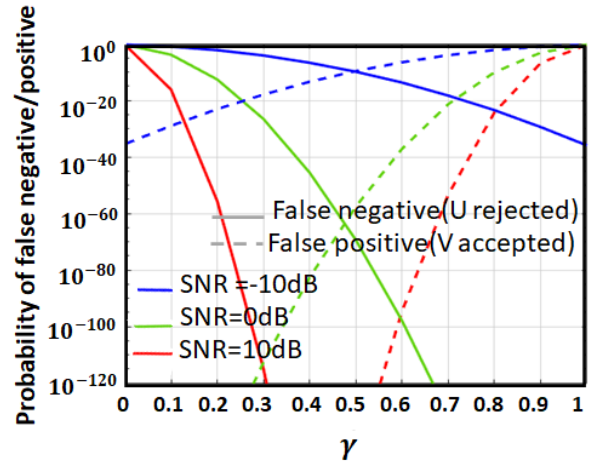

Fig. 9. Probability of false positive and negative as a function of $\gamma$ for different SNR values.

**B. Path identification in optical networks**

Let us consider the application of OI to an optical network. First, we take into account the simple case of two sub-systems (e.g., two optical fiber spans), then we extend our method to a large number of sub-systems in the network. Finally, two examples are considered for specific network topologies: point-to-point in the access segment, and point-to-point in the metro/core infrastructure.

*B.1 Two sub-systems identification*

C-OFDR allows to measure RBPs provided by a single fiber/pigtail or a fiber concatenation. In optical networks, different fiber spans as well as passive sub-systems are plugged through fiber connectors which cause reflections due to the fiber-air interfaces. Such reflections appear as high back-reflected intensity peaks at each specific distance thus giving information about the link composition within the network. A typical reflectivity as a function of the distance including Rayleigh backscattering (RB) and connectors is shown in Figure 10. In the figure, an arbitrary fiber segment is identified for each fiber span through red and blue points respectively.

Network path identification is based on the measure and identification of the two concatenated fiber span signatures provided by segments RBPs. The procedure is detailed hereafter.

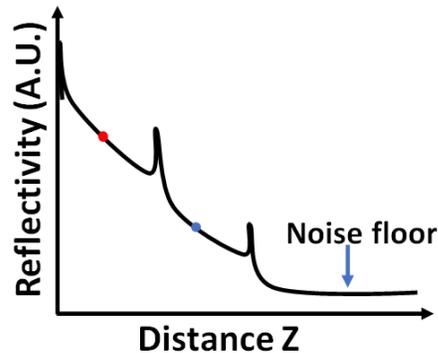

Fig.10 Backscattering typical reflection profile measured across three fiber connectors (peaks) and two fiber spans. Red and blue dots are chosen for fiber identification.

First, the RBPs associated to each fiber are measured through C-OFDR, using the same challenge as is used in section 5. A, and acquired with a high-resolution ADC (6 bit). Next, a scanning window is applied to the RBP of each fiber to select a specific fiber subsection long enough to uniquely identify the fiber. We here take 26 cm for each fiber at the arbitrary distance shown in Figure 10 as red and blue points respectively. RBPs amplitudes are normalized to their maximum values. Figure 11 (top) shows the measured RBPs for the two fibers at the red (left) and blue (right) positions as described in Figure 10. For each RBP a reduced set of data $S_1$, $S_2$ is selected for each fiber (red and blue color in the pictures respectively). $S_1$ and $S_2$ are independently interpolated and overlapped to calculate the intersection points as shown in Figure 11 (bottom). Finally, the intersection points are quantized to two levels to obtain a digital signature, represented as a QR code as in section A. The signature represents the specific ID of the considered network path and is used for path identification through HD with reference IDs. If an additional shared key is added to the signature, this should also be stored in the database.

The protocol is detailed as follows:

Protocol 1: Two sub-systems identification protocol
1. Measure the RBP of the two sub-systems
2. Select RBP for the two sections → $S_1$ and $S_2$
3. Find $S_1$ and $S_2$ intersection points
4. Two levels quantization → ID generation
5. HD evaluation with reference ID → identification

For the performance evaluation of the proposed method, 200 independent network IDs (QR) are generated and stored into the database together with the related keys. Thus, the *challenge* is the sweeping parameters to measure the RBP, and the *response* is the binary image obtained from the RBP of the selected part of the fibers in the networks. Additionally, 25 fake IDs are randomly generated for testing purposes. After HD evaluation, none of the fake 25 IDs are identified as belonging to the database, and only genuine 200 IDs give positive responses each having a single match.

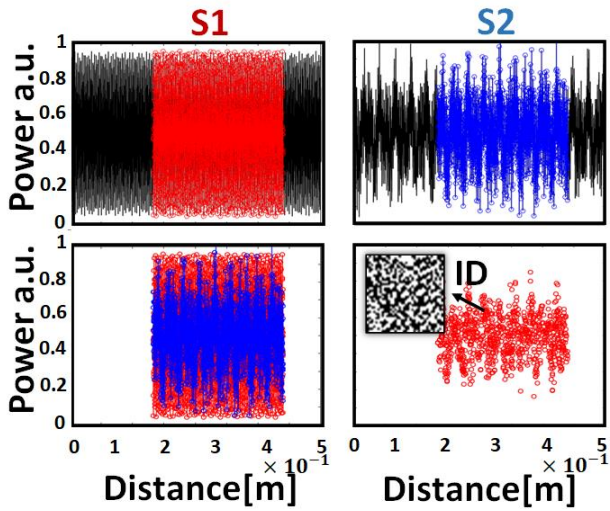

Fig. 11. Top: Selected backscattered data for the two fibers (left and right); bottom: (left) data superimposition, (right) intersection points, and binary image (QR) as network ID.

*B.2 Cascaded sub-systems identification*

ID generation is here described for a system including three devices (fibers) but can be extended to any number of sub-systems. We describe here two possible approaches.

The first approach simply extends what has been described in section B.1 for the two sub-systems by intersecting the measure of the third sub-system $S_3$ with the intersections of the previous two. The binary ID is finally distilled by two-level quantization and is used for identification through HD measure with reference IDs.

Alternatively, the three-fiber ID may be obtained by cascading two independent intersection points of sub-systems measure (e.g., 1, 2 and 1, 3). The resulting ID is obtained as a subset of the concatenation of intersection points after two-level quantization. Identification is obtained by HD evaluation with reference IDs. The protocols for both approaches are listed hereafter (Protocol 2,3).

Protocol 2: Cascaded sub-systems identification protocol (I)
1. Measure the RBP of the cascaded sub-systems
2. Select RBP for three sections → $S_1$, $S_2$, $S_3$
3. Find $S_1$ and $S_2$ intersection points → $J_1$
4. Find $J_1$ and $S_3$ intersection points → $J_2$
5. $J_2$ two levels quantization → ID generation
6. HD evaluation with reference ID → identification

Protocol 3: Cascaded sub-systems identification protocol (II)
1. Measure the RBP of the cascaded sub-systems
2. Select RBP for three sections → $S_1$, $S_2$, $S_3$
3. Find $S_1$ and $S_2$ intersection points → $J_1$
4. Find $S_1$ and $S_3$ intersection points → $J_2$
5. Take a subset $(J_1, J_2) → J_3$
6. $J_3$ two levels quantization → ID generation
7. HD evaluation with reference ID → identification

*B.3 Point-to-point identification in the access segment*

In an optical network, if the transmitter (Tx) and/or receiver (Rx) are able to perform RBP measurement, the identification will be implemented through direct measurement and HD evaluation. In an access network architecture as shown in Figure 12, the central node (CN) communicates point-to-point (P2P) with any connected user. The proposed technique allows to have mutual identification between CN and users if both have access to the database where ID are stored. In Figure 12 an identification example is shown. The CN sends challenge $C_1$ through the fiber link to the user with $ID_1$ and collects back response $R_1$ represented by the RBP in black in the figure. A subset of the data is extracted (blue curve), sampled (red curve) and digitized to obtain the ID (QR). The CN, in this case, searches for ID into the database, and, if it finds a match, it confirms user identification. A symmetric procedure is adopted by the user for CN identification.

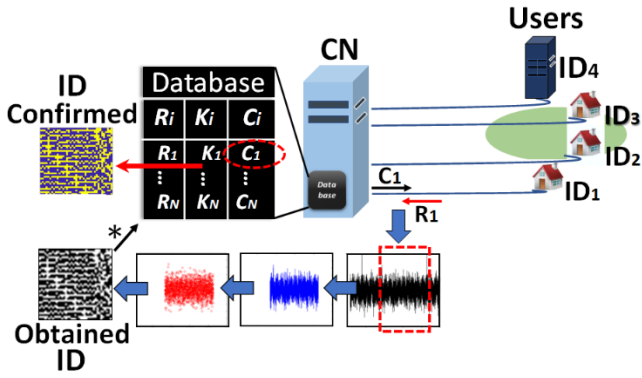

Fig. 12. User $ID_1$ identification procedure by central node (CN) in a P2P architecture.

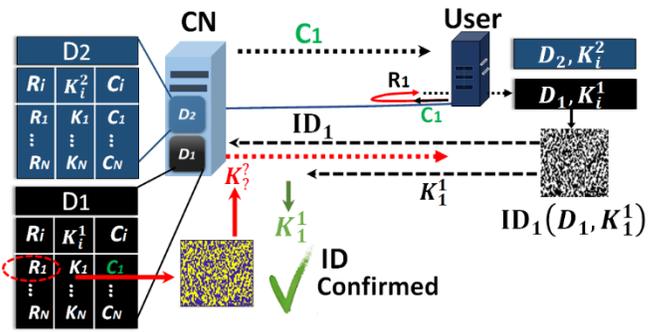

Fig. 13. Central node (CN) assisted user identification in a metro/core infrastructure.

### B.4 Point-to-point identification example in a metro/core infrastructure

In metro and core optical networks, where paths may overcome 50km length, RBP measures can be very noisy or not feasible due to fiber attenuation. We propose here a central node (CN)-assisted identification strategy for long distances exploiting RBP detailed in Figure 13. Two ID databases $D_1$, $D_2$ include the same set of challenges and responses and a specific key $k_1$, $k_2$. The CN communicates to the user a specific challenge $C_1$ that the user must send for link identification measure in the proximity of his site (fiber pigtail of last mile fiber). The response $R_1$ is collected by the user as the RBP and ID is obtained though concatenation with one of the two keys arbitrarily (e.g., $k_1$). The ID is sent to the CN who searches for a match into both $D_1$ and $D_2$. If the CN finds the ID into one of the two databases, it asks the user which of the two keys he did use. If the key corresponds to the database where the match was found, the ID is confirmed. In this way, the access to the link by an adversary that may read the RBP, is not enough for the identification process that is protected by the knowledge of the used key. The identification protocol is provided below (protocol 4).

In contrast to the traditional PUF challenge-response database, in which any single challenge ends with a response, in the proposed method, each challenge can be used for several fibers. That is, one challenge could be used for several fibers providing different responses. This method also reduces the OPUF challenge-response database size.

Protocol 4: Point-to-point identification protocol in a metro/core infrastructure

1. CN sends the challenge $C_1$ to the user
2. The user measures response $R_1$
3. $R_1$ two levels quantization
4. Arbitrary key concatenation→ ID generation
5. The user sends ID to the CN
6. CN searches for ID matches in the two databases $D_1$ and $D_2$
   a. Match is found
      i. CN asks the user for the key
      ii. CN verifies the key-database correspondence → identification
   b. Match is not found → identification failed

## 6. CONCLUSION

In this work, we have proposed for the first time the concept of optical identification (OI) for network security purposes at the physical layer. In particular, we propose to use the fingerprint of each network's sub-system to identify, authenticate, and monitor an optical network. We introduce the concept of OI and we describe a technique to assess the identification reliability. Next, we propose to use the physical unclonable functions (PUFs) for OI, to exploit the intrinsic unclonability and uniqueness of PUFs. In particular, we propose to use the Rayleigh backscattering, an optical PUF, to identify a fiber. In this way, each sub-system of a network can be identified just using its pigtail (or the fiber itself for a link), without any additional device, and working directly at the physical layer. To highlight the huge possibilities of this technique and its potential impact in the field, we described two possible applications of OI: physical layer security in classical communications and user authentication in QKD. On the one hand, we highlighted how OI at the physical layer can significantly enhance the security of an optical network through the identification of sub-systems and links. On the other hand, we described how OI can be effectively used for the authentication of the users (Alice and Bob) in a QKD system. Indeed, the authentication of the users (sometimes referred to as authentication of the classical channel) before QKD transmission is essential to ensure the reliability of the whole process and is usually done with conventional cryptographic methods (e.g., the MAC).

We described, as practical OI applications in optical communication systems and networks, how OI can be implemented for the identification of a fiber or a fiber pigtail in a point-to-point scenario and for the identification of a path in an optical network with passive components.

The OI concept represents an innovative approach to physical layer security which can be applied to any optical communication system and network. OI is based on existing intrinsic characteristics of physical sub-systems and it provides additional features to optical systems and networks operation.

**Funding Information.** HORIZON-JU-RIA (101096909), HORIZON-RIA (101092766), EU – NGEU (PE00000014), NRRP (PE00000001), NOP on Research and Innovation (2014-2020 (IV.5)